\documentclass[useAMS,usenatbib,letterpaper,onecolumn]{mn2e}
\usepackage[totalwidth=480pt,totalheight=680pt]{geometry}
\usepackage{graphicx,amssymb,amsbsy}
\input psfig

\newcommand{\be}{\begin{equation}}
\newcommand{\ee}{\end{equation}}
\def\lta{\raise 0.3 ex\hbox{$ < $}\kern -0.75 em
 \lower 0.7 ex\hbox{$\sim$}}
\def\gta{\raise 0.3 ex\hbox{$ > $}\kern -0.75 em
 \lower 0.7 ex\hbox{$\sim$}} 
\newcommand{\xhat}{{\hat x}} 
\newcommand{\yhat}{{\hat y}} 
\newcommand{\zhat}{{\hat z}}
\newcommand{\hhat}{{\hat h}}
\newcommand{\khat}{{\hat k}} 
 
\newcommand{\mfont}{\mathbb}
\newcommand{\ewig}{{\cal E}}

\title[On the Stability of Star-Planet-Moon Systems]
{The Stability of Tidal Equilibrium \\
for Hierarchical Star-Planet-Moon Systems} 

\author[Adams \& Bloch]{Fred C. Adams$^{1,2}$ and Anthony M. Bloch$^{3}$\\
$^1$Physics Department, University of Michigan, Ann Arbor, MI 48109\\
$^2$Astronomy Department, University of Michigan, Ann Arbor, MI 48109\\
$^3$Math Department, University of Michigan, Ann Arbor, MI 48109 }

\begin{document} 

\date{June 2016}

\pagerange{\pageref{firstpage}--\pageref{lastpage}} \pubyear{2016}
\maketitle

\label{firstpage}

\begin{abstract}
Motivated by the current search for exomoons, this paper considers the
stability of tidal equilibrium for hierarchical three-body systems
containing a star, a planet, and a moon. In this treatment, the energy
and angular momentum budgets include contributions from the planetary
orbit, lunar orbit, stellar spin, planetary spin, and lunar spin. The
goal is to determine the optimized energy state of the system subject
to the constraint of constant angular momentum.  Due to the lack of a
closed form solution for the full three-body problem, however, we must
use use an approximate description of the orbits. We first consider
the Keplerian limit and find that the critical energy states are
saddle points, rather than minima, so that these hierarchical systems
have no stable tidal equilibrium states. We then generalize the
calculation so that the lunar orbit is described by a time-averaged
version of the circular restricted three-body problem. In this latter
case, the critical energy state is a shallow minimum, so that a tidal
equilibrium state exists. In both cases, however, the lunar orbit for
the critical point lies outside the boundary (roughly half the Hill
radius) where (previous) numerical simulations indicate dynamical
instability. These results suggest that star-planet-moon systems have
no viable long-term stable states analogous to those found for
two-body systems.
\end{abstract}


\section{Introduction} 
\label{sec:intro} 

A classic dynamical problem is to consider the tidal equilibrium
states for self-gravitating systems that include both rotational and
orbital motion \citep{darwin1,darwin2}.  The conditions required for
the existence of such equilibrium states has been determined
previously for binary star systems \citep{counselman,hut1980}. With
the relatively recent discovery of exoplanets, this problem has
received renewed interest \citep{levrard,soko,abtide}. In this
contribution, we extend previous treatments to include the presence of
a satellite or moon orbiting the secondary (see also
\citealt{barnes,scheeres}). We thus consider the existence of tidal
equilibrium states for hierarchical triple systems consisting of a
star, planet, and moon, where all three bodies have spin angular
momentum.

Although the discovery of moons in extrasolar planetary systems has
not yet been realized, they have generated great interest. Now that
the existence of exoplanets is well established, and their populations
are being characterized, the next astronomical frontier is to discover
moons in other solar systems. These additional bodies are in principle
observable through their transit timing variations \citep{kippingttv},
wherein additional bodies in a planetary system change the times at
which the planet casts shadows on the host star \citep{agol,holman}.
These moons are most readily detected if they have large relative
masses and their host planets orbit near the star. Another regime of
interest is that of potentially habitable moons, where the orbit of
the host planet resides within the habitable zone \citep{kasting} of
its star. In this case, a moon orbiting the planet could be
potentially habitable provided that it has the proper mass,
atmosphere, and other characteristics \citep{kippinghab,heller}. In
both regimes, however, moons are susceptible to removal
\citep{donnison,weidner,spalding}.

In general, systems evolve toward lower energy states, but are
required to conserve their total angular momentum (in the absence of
external torques). For two-body systems, previous treatments show that
three evolutionary paths are available: [A] The orbit of the secondary
can decay inward and eventually collide with the primary, where the
orbital angular momentum is transferred to the rotation of the
primary. [B] The orbit can gain angular momentum from the primary and
move outwards toward an unbound state. [C] The system can approach a
stable tidal equilibrium configuration, where the orbit and spins of
both bodies have the same period and their angular momentum vectors
point in the same direction. The Pluto-Charon system provides one such
example \citep{tholen}. In order for the equilibrium state to exist,
the system must have a minimum amount of total angular momentum; in
order for the system to reside in the equilibrium state, its orbital
angular momentum must be at least three times larger than the spin
angular momentum. The goal of this paper is to derive analogous
requirements for the existence of stable tidal equilibrium states for
hierarchical three-body systems (star-planet-moon systems).

The existence of tidal equilibrium states does not depend directly on
the energy dissipation mechanisms that allow systems to attain such
states. In the cases of interest, the relevant dissipative evolution
is generally driven by tidal effects \citep{hut1981,zahn}. Since tidal
evolution occurs over long time scales, often comparable to the age of
the universe, many extant systems are not expected to have reached
their lowest energy states.  Notice also that tidal equilibrium is
determined under the assumption of conservation of angular momentum,
which must exceed a minimum value for the equilibrium state to
exist. Although planetary systems can lose angular momentum via
stellar winds and other astronomical processes, systems that start
with too little angular momentum generally have no means of gaining
more.

In the regimes of interest, the host planet often has mass comparable
to Jupiter and the moon has mass comparable to Earth. This satellite
mass is favored because it is large enough to produce measurable
transit timing variations in Hot Jupiter systems and large enough to
be potentially habitable in systems with more temperate Jupiters.
Here we denote the masses of the three bodies as $M$ for the star, $m$
for the planet, and $\mu$ for the moon. These masses are assumed to
obey the ordering 
\be\mu\ll{m}\ll{M}.\label{ordering}\ee
This work assumes that the planetary orbit (about the star) has
semimajor axis $a$ and the moon orbit (about the planet) has semimajor
axis $b$ (when the moon orbits within a non-Keplerian potential, the
scale $b$ refers to the radius of the lunar orbit). The star has spin
angular momentum with moment of inertia $I$ and rotation rate
$\Omega$.  The planet also has spin angular momentum with moment of
inertia $J$ and rotation rate $\omega$.  Finally, the moon has moment
of inertia $K$ and rotation rate $\lambda$.  Because of the mass
ordering from equation (\ref{ordering}), we can ignore the difference
between the reduced mass and the actual mass of an orbiting body.

A great deal of previous work has considered the dynamics of these
hierarchical systems \citep{szebehely,szebehely78}. However, most of
the work regarding system stability does not include the effects of
rotation of the bodies (see the review of \citealt{george}). As shown
previously for two-body systems, the possible exchange of angular
momentum between the orbit(s) and rotation plays an important role in
determining tidal equilibrium states. We also note that many previous
numerical simulations (e.g., \citealt{payne} and references therein)
have shown that the long-term dynamical stability of the system
requires the lunar orbit to fall within some fraction $f$ of the Hill
radius $(b<fR_H)$, where $R_H\equiv(m/3M)^{1/3}a$ and $f\sim1/2$.
These numerical studies generally do not include rotational angular
momentum, but provide independent constraints on the stability of
star-planet-moon systems. Significantly, the results of this paper
indicate that tidal equilibrium states --- when they exist ---
correspond to lunar orbits that lie outside the stability regime found
numerically. Finally, the tidal evolution of our moon, and others,
including tidal dissipation has been studied in detail
\citep{goldpeale,touma}). The dissipation time scales often exceed the
age of the system, or even the age of the universe, so that observed
systems are often not found in their lowest accessible energy states.

The goal of this paper is to consider the equilibrium states of the
system, including the effects of stellar, planetary, and lunar
rotation in the energy and angular momentum budgets. We first consider
the problem in the limit where both the orbits can be described using
Keplerian solutions (Section \ref{sec:kepler}). For this case, we
allow the five angular momentum vectors to have different directions
and allow the orbits to have eccentricity.  Nevertheless, the tidal
equilibrium state corresponds to circular orbits and aligned angular
momentum vectors.  Moreover, the critical point is a saddle point,
rather than a minimum, so that no stable tidal equilibrium state
exists for this limiting case. We then generalize the problem to
include the stellar influence on the lunar orbit, which is treated by
a time-average of the circular restricted three-body problem (Section
\ref{sec:beyond}). For this case, the system has two critical points.
The first corresponds to a saddle point, whereas the second
corresponds to a minimum of the energy so that a stable tidal
equilibrium state exists. In Section \ref{sec:apply}, we briefly
consider applications of these results for extrasolar planetary
systems, as well as moons in our Solar System.  The paper concludes,
in Section \ref{sec:conclude}, with a summary of our results and a
discussion of their implications.

\section{Three-Body Systems in the Keplerian Limit} 
\label{sec:kepler} 

As a first approximation, we consider the orbits of both the planet
and the moon to be described by standard Keplerian solutions. The
energy and angular momentum of the system thus have five
contributions: the planetary orbit around the star, the lunar orbit
around the planet, stellar spin, planetary spin, and lunar spin.
Because of the ordering of masses from equation (\ref{ordering}), the
center of mass of the system lies at the center of the star.

In physical units, the energy and anuglar momentum of the system are given by 
\be{E}=-{GMm\over2a}-{Gm\mu\over2b}+{1\over2}I|{\vec\Omega}|^2+
{1\over2}J|{\vec\omega}|^2+{1\over2}K|{\vec\lambda}|^2.\label{energy}\ee
The first two terms arise from the energy of the two orbits 
and the next three terms arise from the rotational energy. 
The total angular momentum $\vec{L}$ is given by 
\be\vec{L}={\vec{h}}+{\vec{k}}+I{\vec\Omega}+J{\vec\omega}+K{\vec\lambda}.\label{angmom}\ee
The orbital angular momenta of the planet $\vec{h}$ and moon $\vec{k}$ 
have magnitudes defined by 
\be{h}\equiv{m}\sqrt{GMa(1-e^2)}\qquad{\rm and}\qquad{k}\equiv\mu\sqrt{Gmb(1-\epsilon^2)},\ee
where $e$ and $\epsilon$ are the eccentricities for the planetary and
lunar orbit, respectively.  In general, the five angular momentum
vectors can have different directions.  We choose the coordinate axes
so that the total angular momentum vector of the system points in the
$\zhat$ direction.  The orbital angular momentum vector for the
planetary orbit has direction given by $\hhat\cdot\zhat=\cos{i}$. 
Without loss of generality, we let the vector lie in the $x$-$z$
plane. Similarly, the angular momentum vector of the lunar orbit has 
direction given by $\khat\cdot\zhat=\cos\theta$, where we introduce 
the azimuthal angle $\phi$ such that $\khat\cdot\xhat=\sin\theta\cos\phi$
and $\khat\cdot\yhat=\sin\theta\sin\phi$.  With these definitions, the
total angular momentum has $z$-component
\be{L}=|{\vec{L}}|=L_z=h\cos{i}+k\cos\theta+I\Omega_z+J\omega_z+K\lambda_z.\label{lz}\ee
The remaining components of the angular momentum vector
must vanish so that 
\be{L_x}=h\sin{i}+k\sin\theta\cos\phi+I\Omega_x+J\omega_x+K\lambda_x=0,\label{lx}\ee
and
\be{L_y}=k\sin\theta\sin\phi+I\Omega_y+J\omega_y+K\lambda_y=0.\label{ly}\ee 

\subsection{Equilibrium States of the System: First Variation} 
\label{sec:firstvar}  

To simplify the calculation, we work in dimensionless units where
$G=M=I=1$, so that the energy of the system takes the form
\be{E}=-{m\over2a}-{\mu m\over2b}+{1\over2}\left(\Omega_x^2+\Omega_y^2+\Omega_z^2\right) 
+{1\over2}J\left(\omega_x^2+\omega_y^2+\omega_z^2\right)
+{1\over2}K\left(\lambda_x^2+\lambda_y^2+\lambda_z^2\right),\ee
where the angular velocity vectors are written in terms of their
components.  Using conservation of angular momentum, we eliminate the
stellar spin variables $(\Omega_x,\Omega_y,\Omega_z)$ from the energy
expression so it becomes 
\be{E}=-{m\over2a}-{\mu m\over2b}+{1\over2}(L-h\cos{i}-k\cos\theta 
-J\omega_z-K\lambda_z)^2+{1\over2}(h\sin{i}+k\sin\theta\cos\phi+J\omega_x+K\lambda_x)^2\ee
$$+{1\over2}(k\sin\theta\sin\phi+J\omega_y+K\lambda_y)^2 
+{J\over2}(\omega_x^2+\omega_y^2+\omega_z^2)
+{K\over2}(\lambda_x^2+\lambda_y^2+\lambda_z^2).$$

In this form, the energy is a function of 13 variables
$(a,b,e,\epsilon,i,\theta,\phi,\omega_x,\omega_y,\omega_z,\lambda_x,\lambda_y,\lambda_z)$. 
The critical point is determined by the condition that the
deriviatives with respect to all 13 variables must vanish. 
These conditions take the forms
\be{\partial{E}\over\partial{a}}={m\over2a^2}-
{h\over2a}\left[\Omega_z\cos{i}+\Omega_x\sin{i}\right]=0,\label{first}\ee
\be{\partial{E}\over\partial{b}}={m\mu\over2b^2}-{k\over2b}
\left[\Omega_z\cos\theta+\Omega_x\sin\theta\cos\phi+\Omega_y\sin\theta\sin\phi\right]=0,\ee
\be{\partial{E}\over\partial{e}}=2m\sqrt{a}\left[\cos{i}\Omega_z+\sin{i}\Omega_x\right]
{e\over\sqrt{1-e^2}}=0,\ee
\be{\partial{E}\over\partial\epsilon}=2\mu\sqrt{mb}
\left[\Omega_z\cos\theta+\Omega_x\sin\theta\cos\phi+\Omega_y\sin\theta\sin\phi\right]
{\epsilon\over\sqrt{1-\epsilon^2}}=0,\ee
\be{\partial{E}\over\partial{i}}=h\left[\Omega_z\sin{i}-\Omega_x\cos{i}\right]=0,\ee
\be{\partial{E}\over\partial\theta}=k\left[\Omega_z\sin\theta-
\Omega_x\cos\theta\cos\phi-\Omega_y\cos\theta\sin\phi\right]=0,\ee
\be{\partial{E}\over\partial\phi}=k\sin\theta\left[\Omega_x\sin\phi-\Omega_y\cos\phi\right]=0,\ee
\be{\partial{E}\over\partial\omega_z}=J(\omega_z-\Omega_z)=0,\ee
\be{\partial{E}\over\partial\omega_x}=J(\omega_x-\Omega_x)=0,\ee
\be{\partial{E}\over\partial\omega_y}=J(\omega_y-\Omega_y)=0,\ee 
\be{\partial{E}\over\partial\lambda_z}=K(\lambda_z-\Omega_z)=0,\ee
\be{\partial{E}\over\partial\lambda_x}=K(\lambda_x-\Omega_x)=0,\ee
and finally 
\be{\partial{E}\over\partial\lambda_y}=K(\lambda_y-\Omega_y)=0.\label{last}\ee  
Note that we use the variables $(\Omega_x,\Omega_y,\Omega_z)$ in 
the above expressions for convenience, where these quantities are 
defined via equation (\ref{lz}--\ref{ly}). 

We need to find the point where equations (\ref{first}--\ref{last})
are simultaneously satisifed.  Following the same arguments given in
previous treatments \citep{hut1980,abtide}, we find that the critical
point corresponds to the conditions 
\be{e}=\epsilon=\omega_x=\omega_y=\lambda_x=\lambda_y=\theta=i=\phi=0 
\qquad {\rm and} \qquad 
\omega_z=\lambda_z=\left({1\over{a^3}}\right)^{1/2}=\left({m\over{b^3}}\right)^{1/2}.\ee  
The latter two quantities are the mean motion of the planetary orbit 
$N=(GM/a^3)^{1/2}$ and the mean motion of the lunar orbit 
$\eta=(Gm/b^3)^{1/2}$ (written here in physical units). Although we
have eliminated the variables corresponding to the spin vector of the
star (using conservation of angular momentum), these spin components
can be evaluated to obtain 
\be\Omega_z=\omega_z=\lambda_z\qquad{\rm and}\qquad\Omega_x=\Omega_y=0.\ee 
The critical point thus corresponds to a state where the rotation
rates of all three bodies and the orbits of the planet and the moon
all have the same frequency ($\Omega=\omega=\lambda=N=\eta$). 
Moreover, the orbits and spins are all aligned and the orbits are
circular ($e=0=\epsilon$).

Using the solution found above, we can evaluate the total angular
momentum $L=|\vec{L}|$. Converting back to physical units, $L$ is 
given by the expression 
\be{L}=\left[m(GM)^{2/3}+\mu(Gm)^{2/3}\right]\Omega^{-1/3}+(I+J+K)\Omega,\ee
where $\Omega=\Omega_z=\omega_z=\lambda_z=(GM/a^3)^{1/2}=(Gm/b^3)^{1/2}$.  
This expression has a mininum value, denoted here as $L_X$, which
represents the minimum angular momentum necessary for a tidal
equilibrium state to exist. This minimum can be written in the form 
\be{L_X}={4\over3}\left[3(I+J+K)m^3(GM)^2\right]^{1/4} 
\left[1+\left({\mu^3\over{m}M^2}\right)^{1/3}\right]^{3/4}.\label{lxzero}\ee 
This minimum angular momentum value differs from that found previously
for two-body systems \citep{hut1980} by the second factor in square
brackets and by the larger moment of inertia.  Since these additional
factors are strictly positive, the minimum angular mometum for three
body systems is greater than that for two-body systems, i.e.,
$L_X$(3-body) $>L_X$(2-body). However, the difference between the two
cases is small, with $\Delta{L_X}={\cal O}(\mu{m}^{-1/3}M^{-2/3})\sim10^{-4}$. 
Moreoever, this correction to $L_X$ is roughly comparable to the
orbital angular momentum expected for a moon. Previous work shows that
both Hot Jupiter systems and the inner planets in the systems
discovered by the Kepler mission generally do not have enough total
angular momentum for a stable tidal equilibrium state to exist
\citep{levrard,abtide}. The same deficit is expected to hold if moons
are added to the planets.

\subsection{Stability of the System: Second Variation} 
\label{sec:secondvar} 

To determine if the critical point found in the previous section is a
minimum of the energy, we must consider the second variation.  More 
specifically, we have to evaluate the Hessian matrix \citep{hesse}, 
which is given by 
\be\mfont{H}_{ij}={\partial^2E\over\partial\xi_i\partial\xi_j}\Bigg|_0,\ee 
where the $\xi_i$ are the 13 variables in the problem and where the
derivatives are evaluated at the critical point. The eigenvalues of
the resulting $13\times13$ matrix determine stability. If the Hessian
matrix is positive definite, or equivalently all of its eigenvalues
are positive, then the critical point corrsponds to a local minimum
and hence a stable tidal equilibrium state.

Fortunately, this matrix can be reduced. First we note that the
derivatives for the variable $\phi$ all become zero when evaluated at
the critical point. This behavior arises because the critical point
corresponds to aligned angular momentum vectors, so that the azimuthal
angle plays no role and can be ignored. The remaining 12 variables lead
to a Hessian matrix with the block form 
\be\mfont{H}=\left[\matrix{\mfont{A}&0&0\cr0&\mfont{B}&0\cr0&0&\mfont{C}}\right],\ee
where all of the entries are $4\times4$ matrices. Here, the 
variables are separated into three groups of four and are 
ordered according to $(a,b,\omega_z,\lambda_z)$ for $\mfont{A}$,
$(i,\theta,\omega_x,\lambda_x)$ for $\mfont{B}$, and
$(e,\epsilon,\omega_y,\lambda_y)$ for $\mfont{C}$. 
The three $4\times4$ submatrices have the forms 
$$\,$$
\be\mfont{A}={m\over4}\left[\matrix{
a^{-3}(-3+ma^2)&{\mu}m^{1/2}(ab)^{-1/2}&2Ja^{-1/2}&2K/\sqrt{a}\cr
{\mu}m^{1/2}(ab)^{-1/2}&{\mu}b^{-3}(-3+\mu{b^2})&2\mu{J}(mb)^{-1/2}&2K\mu(mb)^{-1/2}\cr
2Ja^{-1/2}&2{\mu}J(mb)^{-1/2}&4J(1+J)/m&4JK/m\cr
2K/\sqrt{a}&2K\mu(mb)^{-1/2}&4JK/m&4K(K+1)/m}\right],\ee
$$\,$$
\be\mfont{B}=m\left[\matrix{a^{-1}(1+ma^2)&\mu(mab)^{1/2}&a^{1/2}J&K\sqrt{a}\cr
\mu(mab)^{1/2}&{\mu}b^{-1}(1+{\mu}b^2)&{\mu}J(b/m)^{1/2}&\mu{K}(b/m)^{1/2}\cr
a^{1/2}J&\mu{J}(b/m)^{1/2}&J(1+J)/m&JK/m\cr{K}\sqrt{a}&\mu{K}(b/m)^{1/2}&JK/m&K(K+1)/m}\right],\ee 
and 
$\,$ 
\be\mfont{C}=\left[\matrix{m/a&0&0&0\cr0&{\mu}m/b&0&0\cr0&0&J(J+1)&JK\cr0&0&JK&K(K+1)}\right].\ee
Note that we could separate the submatrix $\mfont{C}$ into two
$2\times2$ sub-submatrices corresponding to the variables
$(e,\epsilon)$ and $(\omega_y,\lambda_y)$. 

In this setting, we do not need the actual eigenvalues, but only need
to know if they are positive or not.  We thus use Sylvester's
Criterion \citep{gilbert}, which states that a real-symmetric matrix
is positive definite if and only if all its leading principal minors
are positive. Recall that a positive definite matrix has positive
eigenvalues.

For submatrix $\mfont{A}$, the requirement that the first
minor is positive implies 
\be{ma^2}>3.\label{minor21}\ee
The requirement that the second minor is postive implies 
\be(-3+ma^2)(-3+\mu b^2)>(ma^2)(\mu b^2)\quad\Rightarrow\quad3>ma^2+\mu{b^2}.\label{minor22}\ee
Since equations (\ref{minor21}) and (\ref{minor22}) cannot be
satisfied simultaneously, the submatrix $\mfont{A}$ is not positive
definite, and the critical point is a saddle point rather than a
minimum. As a result, no stable equilibrium state exists for this
system.

For completeness, we note that the first two minors of submatrix
$\mfont{B}$ are positive. The requirement that the third minor is
positive takes the form 
\be1+\mu{b^2}>(\mu{b^2})(ma^2).\ee
This requirement is usually, but not always, satisfied. When this
condition is violated, the matrix has another negative eigenvalue.
(Note that the eigenvalues of the submatrix $\mfont{C}$ are always
positive.)

This result --- that no stable tidal equilibrium state exists ---
stands in contrast to that obtained earlier for two-body systems
\citep{hut1980}, including higher-order terms for the potential
\citep{abtide}. The addition of the moon removes the possibility of
the system attaining tidal equilibrium. The critical point of the
system, which corresponds to synchronous rotation, becomes a saddle
point, rather than a minimum, with the inclusion of the moon. 

\begin{figure} 
\centerline{\includegraphics[width=12.5cm]{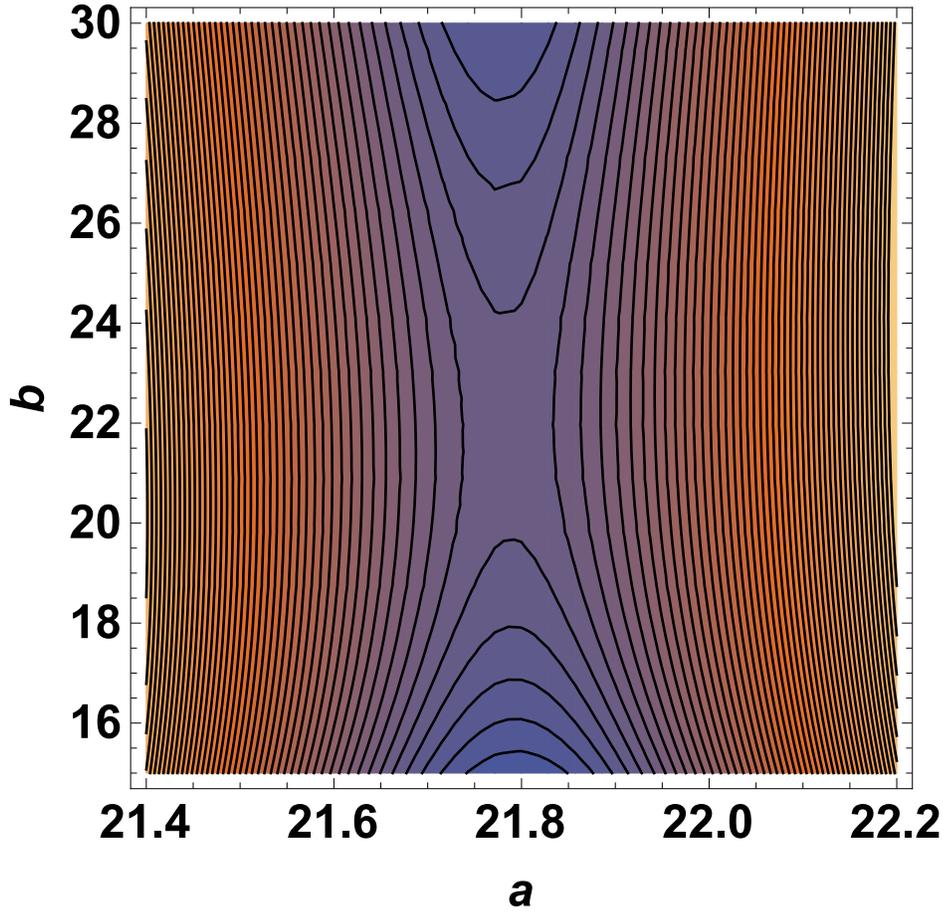}} 
\caption{Energy contours for a reduced system in the Keplerian limit, 
where energy and angular momentum budgets include the planetary orbit,
planetary spin, and lunar orbit. The critical point is a saddle point
and occurs near the center of the diagram. The energy has a local
minimum in the $a$-direction and a local maximum in the
$b$-direction. } 
\label{fig:saddle} 
\end{figure}  

The existence of a saddle point arises because the system can transfer
angular momentum between the lunar orbit and planetary orbit (and the
planetary spin). To illustrate this behavior, we introduce a reduced
system with only the two orbits and the planetary spin. The specific
energy $\ewig=E/m$ takes the form 
\be\ewig=-{1\over2a}-{\delta\over2b}+
{m\over2}(\Lambda-\sqrt{a}-\delta\sqrt{b})^2\,,\label{model}\ee
where we have defined $\delta=\mu{m}^{-1/3}$ and $\Lambda=L/m$, and
let $b\to{m^{-1/3}}b$.  The contours of constant energy are shown in
Figure \ref{fig:saddle} for one such system (where we have taken
$\delta=10^{-3}$, $m=0.03$, and $\Lambda=5$. The energy has a critical
point near the center of the diagram. The point is a minimum in the
direction corresponding to variations in the planetary semimajor axis
$a$, but is a local maximum in the orthogonal direction corresponding
to variations in $b$.

\subsection{Reduction to the Two-body System} 
\label{sec:twobody} 

As a consistency check, we consider the stability analysis in the
limit where the moon mass is negligible by taking $\mu\to0,K\to0$,
where we should recover the results found earlier for binary
systems. In this limit, the three $4\times4$ submatrices of the
Hessian matrix reduce to three $2\times2$ submatrices, with the
variables $(a,\omega_z)$ for $\mfont{A}_2$, $(i,\omega_x)$ for
$\mfont{B}_2$, and $(e,\omega_y)$ for $\mfont{C}_2$: 
\be\mfont{A}_2={m\over4}\left[\matrix{a^{-3}(-3+ma^2)&2Ja^{-1/2}\cr
2Ja^{-1/2}&4J(1+J)/m}\right],\ee
\be\mfont{B}_2=m\left[\matrix{a^{-1}(1+ma^2)&a^{1/2}J\cr{a}^{1/2}J&J(1+J)/m}\right],\ee
and 
\be\mfont{C}_2=\left[\matrix{m/a&0\cr0&J(1+J)}\right].\ee
It is straightforward to show that the eigenvalues of the reduced
submatrices $\mfont{B}_2$ and $\mfont{C}_2$ are always positive. The
conditions required for the reduced submatrix $\mfont{A}_2$ to be
positive definite take the forms 
\be{m}a^2>3\qquad{\rm and}\qquad ma^2>3(1+J).\label{hsthree}\ee
The first condition is satisfied if the second one holds. The second
condition can be rewritten in more suggestive form by converting back
to physical units and rearranging to obtain $m(GMa)^{1/2}>3(I+J)\Omega$. 
In other words, the critical point is a minimum, and a tidal
equilibrium state exists, provided that the orbital angular momentum
is larger than three times the spin angular momentum of the system
(consistent with previous results). Recall that in order for the
critical point to exist, the total angular momentum must be larger
than a minimum value $L_X$, given by equation (\ref{lxzero}). In the
limit $\mu\to0$, this lower bound becomes
\be{L_X}={4\over3}\left[3(I+J)m^3(GM)^2\right]^{1/4}\label{lxone}.\ee 

\section{Minimal Model Beyond the Keplerian Limit} 
\label{sec:beyond}  

This section takes the problem beyond the Keplerian approximation.
The previous treatment did not include the tidal influence of the star
on the lunar orbit or the fact that the lunar orbit resides in a
rotating reference frame. These effects are included here by
considering the circular restricted three-body problem
\citep{szebehely} for the lunar orbit and by using an orbit-averaged
approach \citep{goldreich66}. The results of this section are thus
more useful for applications to real systems.  

In this treatment, both the lunar orbit and the planetary orbit are
taken to have no eccentricity. Motivated by the results of the
previous section, we consider a reduced system where the energy and
angular momentum budgets include only the planetary orbit, the lunar
orbit, and the planetary spin. In particular, we ignore stellar
rotation, so that this calculation applies to planetary orbits with
semimajor axes large enough that the stellar angular momentum is
decoupled. Finally, we assume that the angular momentum vectors of the
two orbits and the planetary spin are aligned. With these
specifications, the energy of the system (in dimensionless units where
$G=M=J=1$) takes the form 
\be 
E = - (m+\mu) {1\over2a} - {m\mu\over2b}
- \mu {b^2 \over 2 a^3} + {1\over2} \omega^2 \,.
\label{energyfull} 
\ee
The quantity $\omega$ = $\omega(a,b)$ is the rotation rate of the
planet and is determined by conservation of angular momentum so that 
\be
\omega \equiv L - (m+\mu) \sqrt{a} - \mu \left[ mb - b^4 /2a^3 \right]^{1/2} \,. 
\label{spinfull} 
\ee 
Equation (\ref{energyfull}) follows from an orbit-averaged approach
\citep{goldreich66} to the circular restricted three-body problem
\citep{szebehely}, whereas equation (\ref{spinfull}) makes the
additional assumption that the orbit of the moon is also circular.
Now we consider the specific energy $\ewig=E/m$ and ignore the
difference between $m$ and $m+\mu$ for the planetary orbit.  We also
define the specific angular momentum $\Lambda=L/m$ and the ancillary
quantity $\delta=\mu m^{-1/3}$.  Finally, we rescale the radial
variable for the lunar orbit $b\to m^{-1/3}b$. The specific energy can
then be written in the form
\be
\ewig = - {1\over2a} - {\delta\over2b} - \delta {b^2\over2a^3} 
+ {m\over2} (\Lambda - \sqrt{a} - \delta \left[b - b^4/2a^3 \right]^{1/2})^2 \,. 
\ee
For future reference note that 
$\omega/m = \Lambda - \sqrt{a} - \delta \left[b - b^4/2a^3 \right]^{1/2}$.

\subsection{Equilibrium States: First Variation} 
\label{sec:fullfirst} 

\noindent 
Now we find the derivatives. The derivative for $a$ has the form
\be
{\partial \ewig \over \partial a} = {1\over2a^2} + 
{3\over2}\delta {b^2 \over{a^4}} - {\omega \over 2} \left\{ a^{-1/2} + \delta
\left[ b - b^4/2a^3 \right]^{-1/2} 3b^4/2a^4 \right\} \,. 
\label{deda} 
\ee 
The derivative for the second semimajor axis $b$ can be written
\be
{\partial \ewig \over \partial b} = {\delta\over2b^2} - \delta{b \over a^3} - 
{\delta \omega \over 2} \left[ b - b^4/2a^3 \right]^{-1/2} 
\left[ 1 - 2b^3/a^3 \right] \,. 
\label{dedb} 
\ee

\noindent 
Setting the derivatives to zero we find 
\be
1 + 3 \delta {b^2 \over{a^2}} =
\omega \left\{ a^{3/2} + \delta 
\left[ b - b^4/2a^3 \right]^{-1/2} 3b^4/2a^2 \right\} \,,
\ee 
and 
\be
\left[ 1 - {2b^3 \over a^3} \right] = \omega b^2 
\left[ b - b^4/2a^3 \right]^{-1/2} 
\left[ 1 - {2b^3 \over a^3} \right] \,.
\ee
The second relation has two solutions. 
The first solution implies 
\be
\omega^2 = {1 \over b^3} - {1 \over 2 a^3} \qquad {\rm or} \qquad 
\eta^2 = \omega^2 + N^2/2 \,.
\label{critcorot} 
\ee
At this critical point, the spin of the planet is synchronous 
with the orbital frequency of the moon (in the rotating frame 
of reference). The second solution implies 
\be
{1 \over b^3} = {2 \over a^3} \qquad {\rm or} \qquad 
\eta^2 = 2 N^2 \,. 
\label{critmaxk} 
\ee
This value of $b$ corresponds to the point where the orbital angular
momentum of the moon has its maximum value. 

The second critical point (where $\eta^2=2N^2$) will generally lie
closer to the planet than the first critical point (where
$\eta^2=\omega^2+N^2/2$). In order for the two critical points to
coincide, the condition $\delta(b/a)^2=(-2+\sqrt{6})/3\approx0.15$
must be met. As shown below, we expect $b/a\approx(2/3)^{1/3}$ at the
critical point.  (Recall that the semimajor axis $b$ is scaled by a
factor of $m^{-1/3}$ so that $a$ and $b$ are comparable in size.) As a
result, this condition implies $\delta \approx 0.2$. However, we
expect $\delta$ to be small, more specifically $\delta=\mu
m^{-1/3}\approx10\mu\approx10^{-4}\ll1$, so the required condition is
usually not satisfied for exomoons, and the first critical point
(equation [\ref{critcorot}]) occurs farther from the planet than the
second one (equation [\ref{critmaxk}]).

\subsection{Stability of the System: Second Variation} 
\label{sec:fullsecond} 

The second derivative for $a$ has the form 
\be
{\partial^2 \ewig \over \partial a^2} = - N^2 - 
6 \delta {b^2 \over a^2} N^2 + {\omega \over 4} N 
+ {\omega \over 2} \delta \left\{ 
\left[ \eta^2 - N^2/2 \right]^{-1/2} 
6 {b^2 \over a^2} N^2 +  
{9 \over 8} \left[ \eta^2 - N^2/2 \right]^{-3/2} 
{b^2 \over a^2} N^4 \right\} 
\ee
$$
+ {m a^2 \over 4} \left\{ N + \delta
\left[ \eta^2 - N^2/2 \right]^{-1/2} 
(3 b^2/2a^2) N^2 \right\}^2 \,,
$$
where we have used the definitions of $N$ and $\eta$
to simplify the expression. The mixed derivative is given by 
\be
{\partial^2 \ewig \over \partial b \partial a} = {3 \delta b \over a} N^2 
+ \delta {\omega \over 2} \left\{ 
-6 \left[ \eta^2 - N^2/2 \right]^{-1/2} 
+ {3 \over 4} \left[ \eta^2 - N^2/2 \right]^{-3/2} 
\left[\eta^2 - 2N^2 \right] \right\} {b \over a} N^2 
\ee 
$$
+ {m \delta \over 4} 
\left[ \eta^2 - N^2/2 \right]^{-1/2} \left[\eta^2 - 2N^2 \right] 
a b
\left\{ N + \delta [\eta^2 - N^2/2]^{-1/2} (3b^2/2a^2) N^2 \right\} \,.
$$
The second derivative with respect to $b$ has the form 
\be
{\partial^2 \ewig \over \partial b^2} = 
- \delta\eta^2 - \delta N^2 + {m \delta^2 b^2 \over 4} 
\left[ \eta^2 - N^2/2 \right]^{-1} 
\left[ \eta^2 - 2N^2 \right]^2 
\ee
$$
+ \delta {\omega \over 2} \left\{ 6 N^2 
\left[ \eta^2 - N^2/2 \right]^{-1/2} 
+ {1 \over 2} 
\left[ \eta^2 - N^2/2 \right]^{-3/2} 
\left[ \eta^2 - 2N^2 \right]^2 \right\} \,. 
$$  

To simplify the expressions for the second derivatives, we 
define the following quantities:
\be
\lambda = \left[\eta^2 - N^2/2\right]^{1/2},  
\qquad A = \eta^2 - 2N^2, \qquad {\rm and} \qquad 
\xi = \delta {b^2 \over a^2} \,.
\ee
The conditions for the first derivatives to vanish then become 
\be
\lambda=\omega \qquad {\rm and/or} \qquad A=0\,, 
\ee
for the derivative with respect to $b$, and 
\be
1 + 3\xi = \omega \left\{ {1\over N} + 
{3 \xi \over 2 \lambda} \right\}\,
\label{acrit} 
\ee
for the derivative with respect to $a$. If we also divide by $N^2$,
the second derivatives simplify as follows: 
\be
\ewig_{aa} = - 1 - 6\xi + {\omega\over4N} 
+ \omega {\xi \over \lambda} 
\left\{ 3 +  {9 \over 16} {N^2 \over \lambda^2} \right\} 
+ {ma^2 \over 4} \left\{ 1 + {3\over2}
\xi {N\over \lambda} \right\}^2 \,,
\ee
\be
\ewig_{ab}= {3 \delta b \over a} 
+ {3\delta \omega \over \lambda} \left\{ -1 
+ {A \over 8 \lambda^2} \right\} {b \over a} 
+ {m\delta ab \over 4} {A \over N\lambda} 
\left\{1 + {3\over2} \xi{N\over\lambda} \right\} \,,
\ee
and 
\be
\ewig_{bb} = - \delta{\eta^2\over N^2} - \delta + 
{m\delta^2 b^2 \over 4} 
{A^2 \over N^2 \lambda^2} + {\delta \omega \over \lambda} 
\left\{ 3 + {A^2 \over 4N^2\lambda^2} \right\} \,. 
\ee 

\subsection{Critical Point for Lunar Orbit with Maximum Angular Momentum} 
\label{sec:critazero}

Now we invoke the critical point condition $\eta^2=2N^2$ so that
$A=0$.  This condition corresponds to the case where the lunar orbit
has its maximum angular momentum.  The second derivatives, when
evaluated at the critical point, represent the elements of the Hessian
matrix, which become: 
\be
\ewig_{aa} = - 1 - 6\xi + {\omega\over4N} 
+ \omega {\xi \over \lambda} 
\left\{ 3 +  {9 \over 16} {N^2 \over \lambda^2} \right\} 
+ {ma^2 \over 4} \left\{ 1 + {3\over2}
\xi {N\over \lambda} \right\}^2 \,,
\ee
\be
\ewig_{ab}= 3 \delta {b \over a} 
\left[1 - {\omega \over \lambda} \right] \,,
\ee
and 
\be
\ewig_{bb} = - 3 \delta 
\left[1 - {\omega \over \lambda} \right] \,.
\ee
To determine if the critical point is a minimum, we evalute the 
minors of the Hessian matrix. Let us define  
\be
\alpha = \ewig_{aa} \qquad {\rm and} \qquad 
\beta = 3 \delta \left[ 1 - {\omega \over \lambda} \right] \,,
\ee
so that 
\be
\ewig_{bb}= -\beta \qquad {\rm and} \qquad 
\ewig_{ab} = {b \over a} \beta\,.
\ee
The conditions for the minors to be positive then take the forms 
\be
\alpha>0 \qquad {\rm and} \qquad 
- \alpha \beta > {b^2 \over a^2} \beta^2 \,.
\ee 
These conditions thus require $\beta<0$ so that 
$\omega>\lambda$. However, since $\eta^2=2N^2$, 
the parameter $\lambda = (3/2)^{1/2}N$ and 
equation (\ref{acrit}) implies that 
\be
{\omega \over \lambda} = \left({2\over3}\right)^{1/2} 
{1 + 3 \xi \over 1 + (3/2)^{1/2} \xi} \,.
\ee
The leading factor is less than unity, whereas the second factor is of
order unity, so that $\omega<\lambda$ and hence $\beta>0$.  In fact,
the right hand side approaches unity only in the limit $\xi\to\infty$,
so that the factor is less than unity for any physical values of the
parameters. As a result, the required condition cannot be satisfied,
the minors cannot both be positive, and the critical point is not a
minimum. 

\subsection{Critical Point for Co-rotating Lunar Orbit} 
\label{sec:critalt} 

We now consider the critical point where $\lambda=\omega$, which
corresponds to the case where the planetary rotation rate and the mean
motion of the lunar orbit are synchronous in the rotating frame of
reference. This condition follows from setting the derivative of the
energy with respect to $b$ equal to zero. The corresponding condition
for the derivative with respect to $a$ implies 
\be
\lambda = N \left[ 1 + {3 \over 2} \xi \right] = f N \,.
\ee
We also have the relations 
\be
\eta^2 = (f^2 + 1/2) N^2 \qquad {\rm and} \qquad 
A = (f^2 - 3/2) N^2 \,. 
\ee
Using these results, the second derivatives (matric elements) become 
\be
\ewig_{aa} = -1+{f\over4} + 3\xi\left\{-1+{3\over16f^2}\right\} + 
{ma^2\over4} \left\{1 + {3\over2}{\xi\over f}\right\}^2 \,,
\ee
\be
\ewig_{ab}= {3\delta b \over a} {f^2-3/2 \over 8f^2}
+ {m\delta \over 4} {(f^2-3/2) ab \over f}
\left\{ 1 + {3\over2} {\xi\over f} 
\right\} \,,
\ee
and 
\be
\ewig_{bb} = \delta (3/2 - f^2) + 
\delta (1 + m\delta b^2) {(f^2-3/2)^2 \over 4f^2} \,.
\ee 

To fix ideas, let's evaluate the matrix elements in the extreme 
limit where $\delta,\xi\to0$. Keeping only leading order terms, 
we find:
\be
\ewig_{aa} = {1\over4} (ma^2-3) \,, \quad
\ewig_{ab}= - {\delta\over16} {b \over a}
(3 + 2ma^2) \,, \quad {\rm and} \quad 
\ewig_{bb} = {9 \delta \over 16} \,.
\ee
The condition for the first minor to be positive is simply 
\be
ma^2 > 3 \,,
\ee
whereas the condition for the second minor to be positive is 
\be
9 (ma^2-3) > {\delta\over4} {b^2 \over a^2}
(3 + 2ma^2)^2\,,
\ee
which can be rewritten in the form 
\be
ma^2 > 3 + \xi \left[ {1\over4} + {1\over3}ma^2 + 
{1\over9} (ma^2)^2 \right]\,.
\ee
Since we expect $\xi\sim\delta\ll1$, the constraint required for the
second minor to be positive is only slightly more restrictive than the
condition required for the first minor to be positive.

Now let's look at the conditions required for this 
critical point to exist. At the critical point, the 
rotation frequency is given by 
\be
\omega^2 = {1 \over b^3} - {1 \over 2a^3} \,,
\ee
which comes from the condition that the $b$-derivative 
vanishes. The requirement that $a$-derivative vanishes 
implies that 
\be
\omega = a^{-3/2} \left[1+{3\over2}\delta{b^2\over a^2}\right]\,,
\ee
and finally we have the definition of $\omega$ in 
terms of the total angular momentum 
\be
\omega = m\left[ \Lambda - a^{1/2} - \delta b^2 \omega \right]\,.
\ee
We thus have three equations for the three unknowns 
$(\omega,a,b)$. We can combine the first two equations, 
by eliminating $\omega$, to obtain 
\be
\left[ 1 + {3\over2} \delta {b^2\over a^2} \right]^2 
= {a^3 \over b^3} - {1 \over 2} \,. 
\ee
Let $x=b/a$ so that we obtain 
\be
{3 \over 2} + 3\delta x^2 + {9\over4}\delta^2 x^4 = {1\over x^3} \,.
\label{xequation} 
\ee
This equation always has one (and only one) positive real 
solution for $x=b/a$ for a given value of the mass parameter $\delta$. 
We can thus take $x$ to be specified. The remaining equation 
implies that 
\be
L=m\Lambda = m a^{1/2} + 
(1+ m\delta x^2 a^2) a^{-3/2} \left[1+{3\over2}\delta x^2\right]\,,
\ee
which can be rearranged to the form 
\be
L = m a^{1/2} \left[1 + \delta x^2 + {3\over2} \delta^2 x^4\right] 
+ \left[1+{3\over2}\delta x^2\right] a^{-3/2} \,,
\ee 
The angular momentum $L$ of the system must exceed a 
minimum value $L_X$ in order for this equation to have 
a real solution, where the minimum is given by 
\be
L_X = {4 \over 3} \left\{ {3 \over m} 
\left[1+{3\over2}\delta x^2\right]
\left[1 + \delta x^2 + {3\over2} \delta^2 x^4\right]^3 
\right\}^{1/4} \,.
\label{lxmin} 
\ee 
In physical units, the minimum value of the angular momentum 
has the form 
\be
L_X = {4 \over 3} \left[3 J G^2 M^2 m^3 
\left(1 + {3\over2}\delta x^2\right) 
\left(1 + \delta x^2 + {3\over2}\delta^2 x^4\right)^3
\right]^{1/4} \,. 
\label{lxminphys} 
\ee
In this expression, the dimensionless composite parameter
$\delta{x^2}=(\mu{b^2})/(ma^2)$, where $a$ and $b$ are now in physical
units, and where the ratio $b/a$ is determined by the solution to
equation (\ref{xequation}). For this minimum value $L_X$, the combined
angular momentum of the orbits (both the planet and moon) is three
times that of the planetary spin. Recall that in the previous case of
two-body systems \citep{hut1980}, the orbital angular momentum of the
binary had to be larger than three times the spin angular momentum for
the stable equilibrium point to exist.  Three-body systems thus
produce a similar result, where the two contributions to the orbital
angular momentum act together in the accounting.

\begin{figure} 
\centerline{\includegraphics[width=12.5cm]{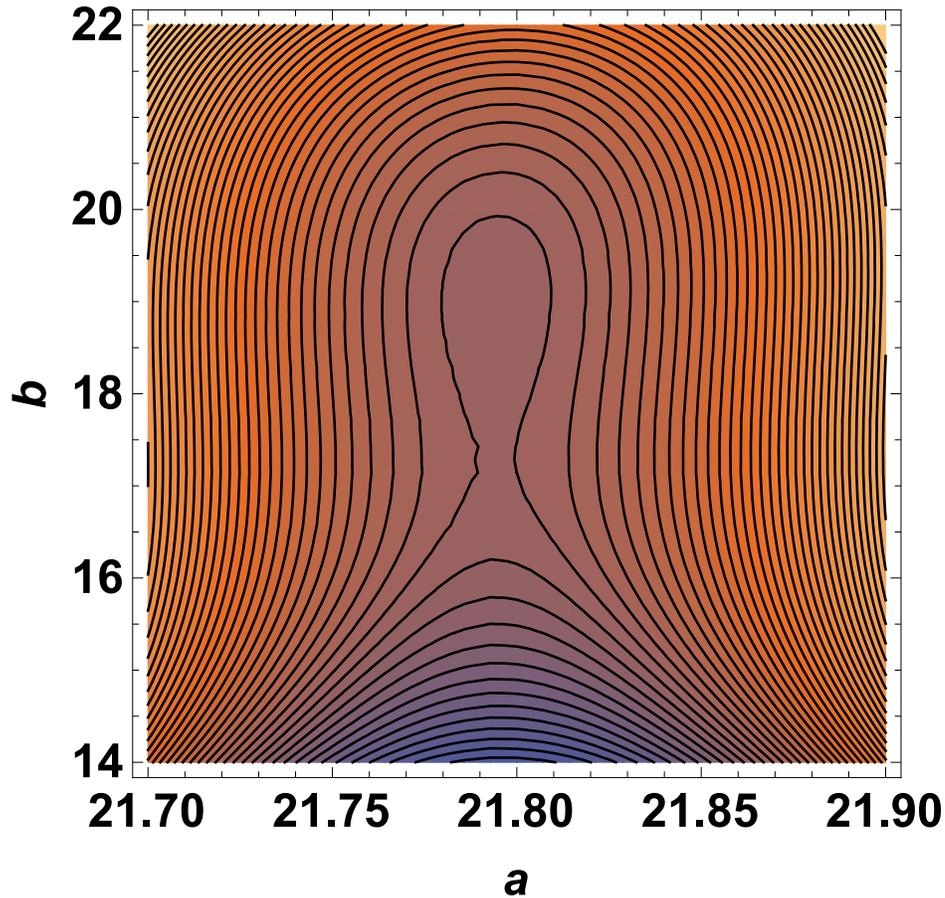}} 
\caption{Energy contours for a reduced system including stellar 
interactions with the lunar orbit. The energy and angular momentum 
budgets include the planetary orbit, planetary spin,
and lunar orbit. The critical point is a minimum, so that a stable 
tidal equilibrium state exists for the system. } 
\label{fig:beyond} 
\end{figure}    

We thus expect this second critical point to be a minimum, provided
that the total angular momentum of the system exceeds that mininmum
given by equation (\ref{lxmin}). Figure \ref{fig:beyond} depicts a
contour plot of the energy for the model system with parameters
$\Lambda=5$, $\delta=10^{-3}$, and $m=0.03$ (the same as those used in
constructing Figure \ref{fig:saddle}). With the generalized treatment
of the lunar orbit, the system has two critical points, where the
first is a saddle point and the second is a minimum. As a result, a
stable tidal equilibrium state exists for the system. 

\begin{figure} 
\centerline{\includegraphics[width=12.5cm]{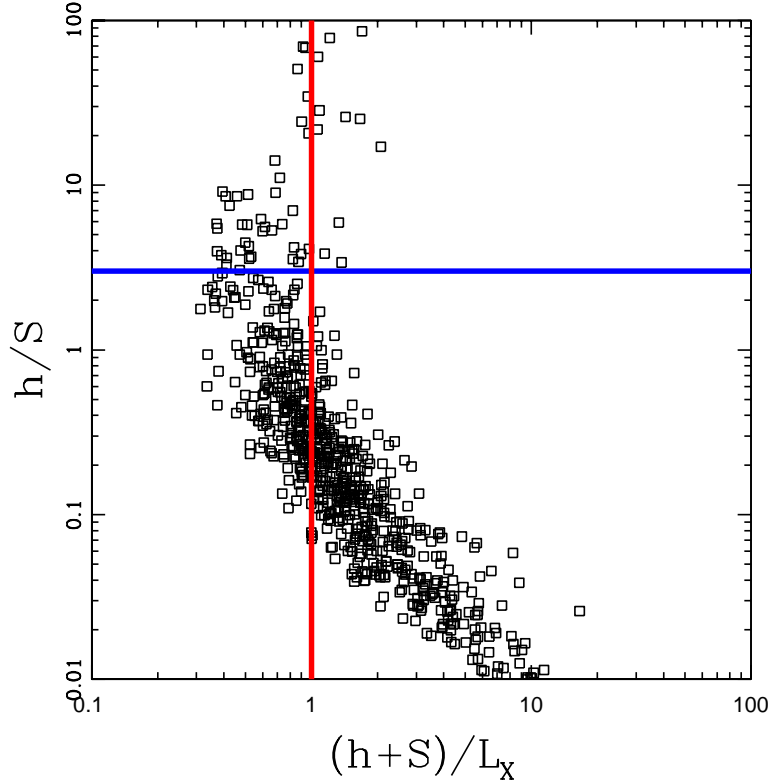}} 
\caption{Angular momentum variables for a collection of observed  
extrasolar planetary systems, where $h$ is the orbital angular
momentum and $S$ is the rotational angular momentum of the star.  
The horizontal axis shows the ratio of total angular momentum to the 
minimum value $L_X$ required for tidal equilibrium states to exist.
The vertical axis shows the ratio of the orbital to spin angular 
momentum. Only systems that fall to the right of the vertical red 
line and above the horizontal blue line can have stable tidal 
equilibrium states.} 
\label{fig:obsplane} 
\end{figure}   

Although a stable tidal equilibrium state can exist, its location is
essentially at the co-rotation point, which is close to the Hill
radius. For comparison, a large body of numerical work suggests that
lunar orbits cannot remain stable over long spans of time if their
orbits are larger than about half the Hill radius (for further
discussion, see \citealt{szebehely78,barnes,domingos,donnison,payne},
and references therein). In other words, lunar orbits corresponding to
the tidal equilibrium point are rendered unstable by dynamical
interactions, so that the tidal equilibrium point found here does not
represent a viable long-term state of the system.

The fact that the tidal equilibrium point occurs near the Hill radius
can be understood as follows: At the Hill radius, the lunar orbit has
nearly the same frequency as the planetary orbit (because the tidal
force due to the star and the gravitational force of the planet act
nearly equally on the moon). Tidal equilibrium for the star-planet
system requires the stellar spin and the planetary spin to coincide
with the orbital frequency of the planet (for further detail, see
\citealt{hut1980}). Similarly, tidal equilibrium for the planet-moon
system requires the planetary spin to coincide with the orbital
frequency of the moon. For the combined star-planet-moon system, in
equilibrium, the orbital frequencies must match up with the spin
frequencies, so that both orbits must have the same period. This
matching of the periods thus leads to the tidal equilibrium point
being near the Hill radius, even though the latter does not involve
rotation rates of the three bodies.

\section{Application to Observed Extrasolar Planetary Systems} 
\label{sec:apply} 

The results of the previous sections outline the properties necessary
for star-planet-moon systems to have stable tidal equilibrium states.
As outlined above, such equilibrium states lie outside the boundary
--- half the Hill radius --- where dynamical scattering is expected to
compromise lunar orbits. Moreover, with no detections of exomoons to
date, we cannot (yet) make a full assessment of the relationship
between exomoon orbits and those corresponding to tidal equilibrium
states in observed extrasolar planetary systems. Nonetheless, existing
data can be used to place constraints.  In this section, we first
consider star-planet systems without moons and show that they
generally cannot have stable tidal equilibrium states. We then find
that hypothetical moons in such systems could have stable tidal
equilibrium states, provided that the star-planet systems have stable
states. Finally, we consider the largest moons in our Solar System and
find that their orbits fall far inside the location of both the tidal
equilibrium state and the dynamical stability boundary.

The results from Section \ref{sec:twobody} outline the properties
necessary for the star-planet system (without a moon) to have a stable
tidal equilibrium state (equations [\ref{hsthree}] and [\ref{lxone}]).
Using these results, consistent with previous findings
\citep{hut1980,abtide}, we can determine whether extrasolar planet
candidates in the current sample meet the criteria for stability.
Here we consider the innermost planets discovered by the {\it Kepler}
mission \citep{batalha}. Since we also need to know the stellar
rotation rates in order to evaluate the stability conditions, we
consider a subset of the sample where the host stars have measured
rotation periods \citep{mcquillan}. After removing eclipsing binaries
and other non-planetary systems, we are left with 738 systems with
detected planets and measured stellar periods. The planets in these
systems primarily have masses in the range $m$ = {\sl few} -- 30
$M_\oplus$ (only 30 planets in the sample have mass $m>0.5m_J$). Note
that we are assuming that the planetary candidates are real and that
their observed radii and inferred orbital elements are accurate.

In order to convert the observed radius measurements $R_p$ to
planetary mass estimates, we use the scaling relationship $m$ =
$M_{\earth} (R_p/R_{\earth})^{2.1}$ \citep{lissauer}. For most of the
sample, the orbital eccentricities are not detectable, and they are
set to zero for this analysis. We also need to specify the moment of
inertia for the host stars, which can be written in the form $I=\chi
M_\ast R_\ast^2$. For simplicity we take $\chi=0.10$, a value that is
intermediate between polytopes with indices $n=3/2$ and $n=3$ (e.g.,
see \citealt{batadams}). The benchmark value of the angular momentum
depends on the moment of inertia according to $L_X\propto I^{1/4}$, so
that the results do not depend sensitively on this choice. 

Subject to the assumptions outlined above, we calculate the orbital
angular momentum $h=m(GMa)^{1/2}$ of the planet and the spin angular
momentum $S=I\Omega$ of the host star.  Figure \ref{fig:obsplane}
shows the ratio of $h/S$ plotted versus the ratio $(h+S)/L_X$, where
$L_X$ is the minimum angular momentum for two-body systems from
equation (\ref{lxone}).  In order for any tidal equilibrium state to
exist for the star-planet system, the total angular momentum must
exceed the critical value $L_X$.  This condition $h+S>L_X$ is shown by
the vertical red line in the figure. In order for the tidal
equilibrium state to be a minimum, and hence stable, the orbital
angular momentum must be larger than three times the spin angular
momentum. This condition, $h/S=3$, is depicted by the horizontal blue
line in the figure. Only those systems that satisfy both criteria can
have stable tidal equilibrium states. As shown by the points in the
upper right quadrant of Figure \ref{fig:obsplane}, only 11 (out of the
original 738) systems meet both stability conditions. The majority of
systems either have no tidal equilbrium state or do not have enough
orbital angular momentum for the equilibrium state to be stable (a
sizable fraction of the systems fail to meet either constraint).  

Most of these systems are not old enough to have reached a stable
tidal equilibrium state (if it exists) or to have lost their planet
(if it does not). Of the 738 planets in the sample, only 239 have
semimajor axes $a<0.05$ AU (and 491 have $a<0.10$ AU), so that the
majority of the planets are effectively decoupled from their stars.
As a result, the time scales for tidal evolution \citep{hut1981,zahn}
are longer than the ages of the systems \citep{basri}.  

We now consider hypothetical moons in orbit about the planets in the
sample. The results from Section \ref{sec:beyond} consider the
star-planet-moon system in the absence of stellar rotation and find
the angular minimum angular momentum $L_X$ necessary for a tidal
equilibrium state to exist (see equation [\ref{lxminphys}]). Keep in
mind that this value of $L_X$ for the star-planet-moon system is not
the same as the minimum value necessary for the two-body system.
Since the total angular momentum must be larger than $L_X$, a
sufficient condition is for the planetary orbital angular momentum to
exceed this value.  We thus require 
\be
m (GMa)^{1/2} > L_X = {4 \over 3} \left[3 J G^2 M^2 m^3 
\left(1 + {3\over2}\delta x^2\right) 
\left(1 + \delta x^2 + {3\over2}\delta^2 x^4\right)^3
\right]^{1/4} \,,  
\ee
which reduces to the constraint 
\be
ma^2 > {256 J \over 27} \left(1 + {3\over2}\delta x^2\right) 
\left(1 + \delta x^2 + {3\over2}\delta^2 x^4\right)^3\approx
{256 \over 27} m R_p^2 \,,
\ee
where $R_p$ is the radius of the planet.  In obtaining the final
approximate constraint, we assume that the parameter combination
$\delta x^2={\cal O}(\mu/M)$ is small. We thus obtain the (sufficient)
condition $a\gta3.08R_p$.  Since planets are usually smaller than
their host stars by an order of magnitude ($R_\ast\sim10R_p$) and the
planetary orbit must lie outside the star ($a>R_\ast$) this condition
is always satisfied. As a result, the star-planet-moon systems will
(almost always) have stable tidal equilibrium states, provided that
the star-planet subsystems have stable tidal equilibrium states. Of
course, as discussed previously, moons orbiting at the equilibrium
locations are susceptible to dynamical scattering.

\begin{figure} 
\centerline{\includegraphics[width=12.5cm]{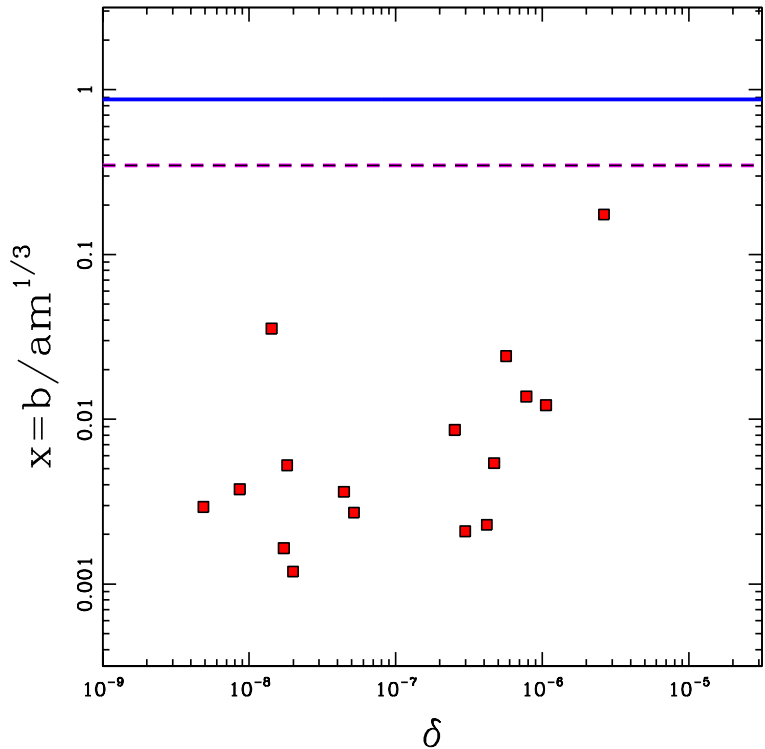}} 
\caption{Scaled orbital ratios for the largest moons in our Solar  
System, plotted versus the scaled mass parameter $\delta$. The upper
blue curve marks the location of the stable tidal equilibrium point. 
The dashed magenta line marks the location of half the Hill radius,
which represents the widest orbit that is expected to be stable
according to previous numerical simulations. The points correspond 
to the 16 largest moons in our Solar System. }  
\label{fig:ssmoon} 
\end{figure}   

Next we apply the results of this analysis to the moons of our own
Solar System. Here we consider the 16 largest moons of the Solar
System.\footnote{The data were take from the Jet Propulsion Laboratory
  website, Planetary Satellite Physical Parameters, ssd.jpl.nasa.gov,
  which includes additional references.} These bodies orbit Earth (our
moon), Pluto (Charon), and the four giant planets.  For a given moon
mass $\mu$ and planetary mass $m$, we calculate the mass parameter
$\delta = \mu/m^{1/3}$, where the masses have been scaled by the mass
of Sun. The solution to equation (\ref{xequation}) then determines the
location of the stable tidal equilibrium point for the moon in terms
of the ratio $b/(am^{1/3})$. This orbital ratio can then be compared
to those of observed moons.

Figure \ref{fig:ssmoon} shows the observed values (red points) of the
scaled orbital ratio as a function of the mass parameter $\delta$ for
the largest moons in our Solar System.  The dashed magenta line shows
the location where the moons would orbit at half their Hill radius.
Numerical simulations \citep{payne} indicate that moons with larger
orbits (above the magenta line) are susceptible to being dynamically
removed from their host planets, so that we do not expect any points
to fall above this line. For comparison, the solid blue curve shows
the value of the orbital ratio corresponding to the stable tidal
equilibrium point (from Section \ref{sec:beyond}). Note that the curve
is nearly constant, as expected for the small values of $\delta$
realized within the Solar System. More importantly, the tidal
equilibrium point (solid blue curve) lies outside the stability
boundary (dashed magenta line). This ordering implies that the
conditions for a stable tidal equilibrium state are at odds with the
requirement for dynamical stability indicated by numerical
simulations.

In Figure \ref{fig:ssmoon}, the point with the largest mass parameter
$\delta$ and the largest orbital ratio (in the upper right part of the
diagram) corresponds to our Moon. The other point that stands out in
the figure is the relatively large orbital ratio for the moon Iapetus,
which orbits Saturn. Except our Moon, Iapetus orbits farther from its
host planet than other (large) moons in our system. With the exception
of the Pluto-Charon system, all of the moons have orbital periods that
are larger than the planetary rotation periods. Because the planets
are spinning faster, tidal forces on the moons are expected to move
them outward toward instability (upward in Figure \ref{fig:ssmoon}).
In contrast, Pluto and Charon and mutually tidally locked. The
rotational periods of Pluto and Charon, as well as their orbital
period, are all about 6.4 days. If considered in isolation from the
Sun, the Pluto-Charon system is thus in a stable equilibrium state for
two-body systems (in addition to synchronicity, the total angular
momentum of the system and the faction of orbital angular momentum are
large enough). However, Figure \ref{fig:ssmoon} indicates that the
three-body system Sun-Pluto-Charon is not in a stable tidal
equilibrium state. The system could evolve to a lower energy state by
transferring angular momentum to its orbit about the Sun.

\section{Conclusion}
\label{sec:conclude} 

This paper considers the issue of tidal equilibrium for hierarchical
star-planet-moon systems, which are described by 16 variables that
specify the energy and angular momentum of the stellar spin, planetary
spin, planetary orbit, lunar orbit, and lunar spin. The optimization
procedure is subject to the constraint of constant angular momentum. A
summary of our results is given below, along with a discussion of
their implications.

\subsection{Summary of Results} 
\label{sec:results} 

The first finding of this paper is that stable tidal equilibrium
states do not exist for hierarchical star-planet-moon systems for the
case where orbits are described in the Keplerian limit (Section
\ref{sec:kepler}).  This result stands in contrast to case of binary
systems with spin angular momentum, also evaluated in the Keplerian
limit, where stable tidal equilibrium states are present as long as
the total angular momentum of the system is sufficiently large. In the
Keplerian approximation, the system energy has a single critical point
which corresponds to synchronous rotation of the star, planet, moon,
planetary orbit, and the lunar orbit. In addition, both orbits have
zero eccentricity and all five angular momentum vectors are aligned.
However, the critical point does not represent a minimum of the
energy, but rather a saddle point (Figure \ref{fig:saddle}). Moreover,
at the critical point, the lunar orbit lies outside the Hill radius,
although this boundary is not defined within the Keplerian
approximation.

In practice, this finding indicates that the system can evolve to a
lower energy state --- while conserving total angular momentum --- by
changing the orbital radius of the moon from its critical value. If
the moon moves inward toward the planet, orbital angular momentum of
the lunar orbit decreases, and the spin and/or the orbital angular
momentum of the planet must increase to compensate. If the moon moves
outward and increases its orbital angular momentum, the spin and/or
orbital angular momentum of the planet decrease accordingly.

In the Keplerian limit, the formulation does not include the tidal
force from the star acting on the lunar orbit. This interaction term
specifies the Hill radius, which provides an effective outer boundary
for allowed lunar orbits. One can include this effect in the
formulation, provided that the additional terms are time-averaged. We
have generalized the calculation so that the lunar orbit is treated
using the circular restricted three-body approximation (Section
\ref{sec:beyond}). In this case, the energy of the system has two
critical points. The first is a saddle point, whereas the second one
is a local minimum (Figure \ref{fig:beyond}). As a result, a stable
tidal equilibrium state exists for this system. The existence of the
tidal equilibrium state requires that the total angular momentum of
the system exceeds a well-defined value $L_X$ (given by equation
[\ref{lxmin}]), analogous to results obtained previously for binary
systems. For $L=L_X$, the combined orbital angular momentum of both
orbits is three times the rotational angular momentum of the planet.
Even when it exists, the stable tidal equilibrium point lies outside
the boundary (roughly half the Hill radius) where lunar orbits are
found to be stable in numerical simulations. As a result, the tidal
equilibrium point found here does not does not represent a viable
long-term state of the system.

Although exomoon systems can in principle have stable tidal
equilibrium states, most observed systems with exo-jupiters don't have
the right angular momentum for a stable tidal equilibrium state to
exist even for the star-planet system considered without a moon (as
illustrated by Figure \ref{fig:obsplane}). For host stars with
measured stellar rotation rates \citep{mcquillan}, the total angular
momentum is generally less than the critical value $L<L_X$ and/or the
orbital angular momentum is too small relative to the total (see also
\citealt{levrard,abtide}). Since these star-planet systems do not
reside in tidal equilibrium, existing systems must dissipate energy
and evolve. However, the time scales must be longer than typical
system ages (several Gyr), and this requirement places constraints on
the tidal dissipation parameters. In a similar vein, moons in our
Solar System orbit much closer to their host planets than both the
tidal equilibrium point and the dynamical stability boundary (Figure
\ref{fig:ssmoon}). Although these moons will eventually be exiled as
they gain orbital angular momentum from their host planets, they
remain in orbit because the Solar System is not old enough for this
process to proceed to completion.

\subsection{Discussion} 
\label{sec:discuss} 

The problem of finding tidal equilibrium states for hierarchical
three-body systems is well-defined in principle. However, the
application of these results to astronomical systems requires further
discussion. First we note that although the full three-body problem
has energy and angular momentum integrals \citep{sundman}, the
expressions used here for these conserved quantities are necessarily
approximate. Because the planetary orbit is nearly Keplerian, the key
approximation in this study is the model used to describe the lunar
orbit. In general, the angular momentum of the moon (in orbit about
the planet) is not constant in time. However, in both the Keplerian
limit (Section \ref{sec:kepler}) and the restricted three-body
generalization used here (Section \ref{sec:beyond}), the time-averaged
lunar orbit does have well-defined angular momentum integral (e.g.,
see \citealt{goldreich66}). As a result, the constrained optimization
procedure of this paper is valid in a time-averaged sense. Since moons
are expected to have periods $P\sim1-100d$, whereas evolutionary times
scales are $\sim$Gyr, the time-averaged angular momentum of the lunar
orbit is indeed well-defined for the time scales over which the system
energy is expected to change.

Another issue that arises is the long-term dynamical stability of
star-planet-moon systems. In the Keplerian limit, the critical point
lies outside the Hill radius, and the critical point is a saddle
point, so that no chance of stability arises. In the orbit-averaged
case (Section \ref{sec:beyond}), the potentially stable critical point
lies near the Hill radius that one obtains for the time-averaged
problem, and the critical point can be a minimum.\footnote[2]{For
  completeness, we note that the Hill radius for the time-averaged
  treatment differs from the Hill radius obtained from the standard
  circular restricted three-body treatment by a factor of order unity.
  In the standard case, one obtains $R_h=(m/3M)^{1/3}a$ for the Hill
  radius, whereas in the time-averaged case, the effective Hill radius
  becomes $R_h=(2m/3M)^{1/3}a$.} However, numerical simulations show
that the stability of lunar orbits requires the semimajor axis to be
less than a fraction (typically $f\sim1/2)$ of the standard Hill
radius (\citealt{payne} and references therein). In scaled units
($G=M=J=1$ and $b\to{b}m^{-1/3}$), this requirement becomes
$b/a<1/(2\cdot3^{1/3})$. In contrast, the value of the ratio $b/a$ at
the stable equilibrium point is given by the solution to equation
(\ref{xequation}), which implies $b/a\approx(2/3)^{1/3}$. The two
locations differ by a factor of $2^{4/3}\approx2.52$. The dynamical
stability constraint thus requires the moon to orbit well inside the
critical point found here. Such systems can evolve to lower energy
states by decreasing the lunar semimajor axis and transferring angular
momentum to the planetary orbit and/or the planetary rotational
energy.

Although star-planet-moon systems often have no stable tidal
equilibrium states, the moons in our Solar System exist over Gyr time
scales as they evolve \citep{goldsoter}. Possible moons in other
systems can have shorter lifetimes \citep{barnes}. Tidal forces act to
move the moons outward (inward) if the planetary rotation rate is
larger (smaller) than the orbital mean motion. For systems that are
dynamically stable, moons must orbit well within the Hill sphere.
Since the planetary spin is close to synchronous with the planetary
orbit for systems near the critical point, which is near the Hill
radius, the lunar mean motion will generally be larger than the
planetary spin; in this case, the moon will eventually fall into the
planet. For systems residing far from their critical point, however,
the planet rotation rate could be super-synchronous, so that the lunar
orbit evolves outward and the moon eventually becomes unbound from the
planet. 

We also note that the tidal equilibrium states considered in this
paper are related to spin-orbit resonances. Many of the moons in our
Solar System are in or near a synchronous spin-orbit resonance where
the orbital period of the moon is commensurate with the rotational
period of the planet \citep{md99}. These configurations correspond to
minimum energy states and planet-moon systems are driven toward such
states through the action of dissipative forces. Tidal equilibrium
states are also minimum energy configurations. In practice,
planet-moon systems in spin-orbit resonance undergo small oscillations
about the equilibrium state. In addition, the full description of the
equilibrium state includes additional properties of the system, such
as the quadrupole moments of the bodies (see Chapter 5 of
\citealt{md99}).

This paper has only considered the energy states of these hierarchical
systems, and not the dynamical evolution toward the equilibrium
states. This evolution must take place through dissipative processes
such as tidal interaction between the constituent bodies. We note that
the time scales for such interactions will generally be different for
the star-planet system and the planet-moon system. As a result, one of
the two-body subsystems can reach its tidal equilibrium state while
the three-body system as a whole remains far from its tidal
equilibrium state. In our Solar System, the Sun-Pluto-Charon system
provides one such example.

The results of this paper change our interpretation of star-planet-moon
systems: Lunar orbits in these systems, including our own Solar
System, often have no tidal equilibrium states and cannot be
absolutely stable (and this result stands in contrast to the case of
two-body systems).  When stable tidal equilibrium states exist, the
required lunar orbits are so distant that dynamical interactions
render the systems untenable, so that the moons can be scattered out
of their orbits. If moons orbit close enough to their host planets to
avoid this fate, they must lie well inside any possible stable
equilibrium point, and must evolve through the action of dissipative
forces (perhaps on long time scales).  As a result, lunar orbits can
only persist because the systems are not old enough for them to have
dissipated their energy, or for dynamical interactions to scatter the
moons out of their planet-centric orbits.


\medskip 
\textbf{Acknowledgments:} We thank Konstantin Batygin, Juliette
Becker, Seth Jacobson, Dan Scheeres and Chris Spalding for useful
conversations. We also thank an anonymous referee for useful comments. 


\label{lastpage}

\end{document}